\documentclass[superscriptaddress,twocolumn,secnumarabic,
amssymb,amsmath,nobibnotes,aps,prd,showpacs,nofootinbib]{revtex4}%
\usepackage{graphicx}
\usepackage{epsf}
\usepackage{bm}
\usepackage{amsmath}
\usepackage{amsfonts}
\usepackage{amssymb}
\usepackage{epstopdf}
\usepackage{natbib}
\usepackage{color}%
\providecommand{\U}[1]{\protect\rule{.1in}{.1in}}
\newcommand{\be}{\begin{equation}}
\newcommand{\ee}{\end{equation}}

\newcommand{\mincir}{\raise
-3.truept\hbox{\rlap{\hbox{$\sim$}}\raise4.truept\hbox{$<$}\ }}
\newcommand{\magcir}{\raise
-3.truept\hbox{\rlap{\hbox{$\sim$}}\raise4.truept\hbox{$>$}\ }}

\usepackage{color}

\begin{document}

\title{Effects of neutrino mass hierarchies on dynamical dark energy models}

\author{Weiqiang Yang}
\email{d11102004@mail.dlut.edu.cn}
\affiliation{Department of Physics, Liaoning Normal University, Dalian, 116029, P. R. China}

\author{Rafael C. Nunes}
\email{rcnunes@fisica.ufjf.br}
\affiliation{Departamento de F\'isica, Universidade Federal de Juiz de Fora, 36036-330, Juiz de Fora, MG, Brazil}

\author{Supriya Pan}
\email{span@iiserkol.ac.in}
\affiliation{Department of Physical Sciences, Indian Institute of Science Education and Research$-$Kolkata, Mohanpur$-$741246, West Bengal, India}

\author{David F. Mota}
\email{mota@astro.uio.no}
\affiliation{Institute of Theoretical Astrophysics, University of Oslo, 0315 Oslo, Norway}


%%%%%%%%%%%%%%%%%%%%%%%%%%%%%%%%%%%%%%%%%%%%%%%%%%%%%%%%%%%%%%%%%%%%%%%%%%%%%%%%%
\begin{abstract}

We investigate how three different possibilities of neutrino mass hierarchies, namely normal, inverted, and degenerate,
can affect the observational constraints on three well known dynamical dark energy models, namely the
Chevallier-Polarski-Linder, logarithmic, and the Jassal-Bagla-Padmanabhan parametrizations.
In order to impose the observational constraints on the models, we performed a robust analysis
using Planck 2015 temperature and polarization data, Supernovae type Ia from Joint Light curve analysis,
baryon acoustic oscillations distance measurements, redshift space distortion characterized by $f(z)\sigma_8(z)$ data,
weak gravitational lensing data from Canada-France-Hawaii Telescope Lensing  Survey, and cosmic chronometers data plus the local value
of the Hubble parameter. We find that different neutrino mass hierarchies return similar fit on almost all model parameters
and mildly change the dynamical dark energy properties.
 
\end{abstract}

%%%%%%%%%%%%%%%%%%%%%%%%%%%%%%%%%%%%%%%%%%%%%%%%%%%%%%%%%%%%%%%%%%
\maketitle
%%%%%%%%%%%%%%%%%%%%%%%%%%%%%%%%%%%%%%%%%%%%%%%%%%%%%%%%%%%%%%%%%%

\section{Introduction}

The hot big bang model predicts the existence of a cosmic neutrino background which has not been directly detected, but has indirectly been 
established by using cosmic microwave background (CMB) observations, as well as estimations from the primordial abundances of light elements. 
On the other hand, the phenomena of neutrino  oscillation from several experiments have shown that neutrinos have a very small but non-zero masses,
see \cite{Garcia} for review. For instance, measures from solar neutrino analysis supplemented by KamLAND estimate
$\Delta m^2_{21} \equiv m_2^2 - m_1^2 \simeq 8 \times 10^{-5}$ eV$^2$ \cite{KamLAND1},
and experiments of atmospheric neutrino oscillation by Super-Kamiokande show
$|\Delta m^2_{31}| \equiv |m_3^2 - m_1^2| \simeq 3 \times 10^{-3}$ eV$^2$ \cite{Kamiokande}.
Unfortunately, the current oscillation experiments do not give much information to the absolute scale of neutrino masses,
since the measures of $\Delta m^2_{21} > 0$ and $|\Delta m^2_{31}|$ lead us to two different possible scenarios of mass
hierarchies, which are known as normal hierarchy (NH) and the inverted hierarchy (IH),  characterized respectively 
by the positive and negative sign of the quantity $|\Delta m^2_{31}|$. 
Within the NH scenario, one eigenstate is much heavier and the lower bound 
is constrained to $\sum m_{\nu} = 0.06$ eV \cite{Olive} [at 95\% confidence level (CL)].  In the IH scenario, the two heaviest neutrinos 
are nearly degenerate and the lower bound is $\sum m_{\nu} = 0.10$ eV \cite{Olive} (95\% CL).
On the other hand, from the point of view of the cosmological restrictions on the neutrino mass bound, one may consider one another
phenomenological hierarchy, the so-called degenerate hierarchy (DH), where the masses of the neutrinos are much larger 
than the differences between them, hence all three active neutrinos are considered to share the same mass. Although NH and IH are two real physical possibilities from the particle physics experiments. 
The possibility that all neutrino masses are virtually the same is not completely excluded by future measurements of absolute mass 
\cite{deGouvea:2013onf}. Since the cosmological data do not have sufficient sensitivity to measure individual masses, 
it is reasonable to consider the total neutrino mass as $\sum m_{\nu} \simeq 3 m_1$ eV (considering three active neutrinos), 
and a lower bound as $\sum m_{\nu} = 0$ eV for DH scheme.
For a general discussions about neutrino mass hierarchies we refer to \cite{Qian}, and the works \cite{Bernardis, Huang, Jimenez, Xu, Gerbino:2016ehw, Vagnozzi}  for cosmological constraints on neutrino mass hierarchies.

Massive neutrinos play an important role on the dynamics of the universe 
affecting important cosmological information sources, for instance, the formation of the large scale structure, big bang nucleosynthesis, 
and CMB anisotropies (see \cite{Dolgov,Lesgourgues} for review). Planck collaboration \cite{Ade:2015xua} within base $\Lambda$CDM + $\sum m_{\nu}$ model has constrained the total 
neutrino mass to $\sum m_{\nu} < 0.194$ eV (from CMB alone), and the effective number of neutrino species, 
$N_{\mbox{eff}}$ $= 3.04 \pm 0.33$ at 2$\sigma$ CL.
An extended $\Lambda$CDM + $c_{\mbox{eff}} + c_{vis} + \sum m_{\nu}$ model has provided 
$\sum m_{\nu} < 0.88$ eV at 95$\%$ CL (from CMB alone) 
\cite{Benjamin}. Forecast on neutrinos from CORE space mission are reported in \cite{CORE}.
Additionally, the cosmological consequences of the massive neutrinos have been investigated in the context of $f(R)$ gravity \cite{fR_nu1,fR_nu2},
holographic dark energy \cite{hde_nu, Wang:2016tsz}, scalar field models \cite{sf_nu,sf_nu2,sf_nu3},
coupled dark energy \cite{coupled_de, coupled_de02, Guo:2017hea}. Furthermore, the presence of massive neutrinos can also reconcile the current tension on the local and global Hubble constant measures \cite{Bernal:2016gxb, DiValentino:2016hlg, Archidiacono:2016kkh}.
Also, the neutrino properties have been considered 
on the estimation of the inflationary parameters \cite{Gerbino:2016sgw,Tram,Valentino}. 

The aim of the present work is to investigate how different neutrino mass hierarchical (or the ordering of the neutrino masses) 
scenarios can correlate with the other cosmological parameters in presence of dynamical  dark energy (DE) models. Since there are 
dozens of different dark energy models \cite{cop,luca,mota1,mota2,mota4,mota5,mota6,mota7,mota8, Lixin1, Lixin2, Nunes:2015rea,Nunes:2016aup,Pan:2016jli,deHaro:2016hpl,deHaro:2016cdm, Yang:2013hra, Nunes:2015xsa, Yang:2014gza, yang:2014vza, Nunes:2014qoa,Pan:2012ki,Nunes:2016dlj,Yang:2016evp, Pantazis:2016nky, Sharov:2017iue}, in this work we consider three general parametric models for DE, namely the Chevallier-Polarski-Linder (CPL) parametrization, the logarithmic model, and the Jassal-Bagla-Padmanabhan (JBP) parametric model.
The dynamical models of DE have been recently constrained from other observational perspectives \cite{DE_1,DE_2,DE_3, DE_4}.
The paper is organized as follows. In the next section we introduce the dynamical dark energy models and their perturbation equations.
Section \ref{sec-data} describes the observational data that we employ in our analysis.
In section \ref{results} we discuss the results of our analysis for all models.
Finally, we close our work in section \ref{conclusions} with a short summary of the whole work.

\section{Dynamical Dark energy}
\label{sec-background}

Let us consider a spatially flat Friedmann-Lema\^itre-Robertson-Walker (FLRW) universe
which is filled with photons ($\gamma$), neutrinos ($\nu$), baryons ($b$), dark matter ($dm$),
and dark energy ($de$) fluids. The Friedmann's equations in such a universe can be written as

\begin{eqnarray}
H^2 = \frac{8 \pi G}{3}\, \left( \rho_{\gamma} + \rho_{\nu} + \rho_b + \rho_{dm} + \rho_{de}  \right),\label{Friedmann1}\\
2\dot{H} + 3 H^2= - 8 \pi G\, (p_{\gamma} + p_{\nu} + p_b + p_{dm} + p_{de})
\end{eqnarray}
where $\rho_i$'s and $p_i$'s ($i= \gamma,\, \nu,\, b,\, dm,\, de$) are respectively the energy density and the pressure of the $i^{th}$ component of the fluid. Moreover, we also consider that the fluid components do not interact with each other. In other words, each
%Here, each
component is conserved separately, that means the balance equation reads

\begin{eqnarray}\label{balance}
\dot{\rho}_i + 3 H (p_i + \rho_i) = 0.
\end{eqnarray}
Now, if the dark energy is of dynamical nature, then its evolution is governed by the following equation

\begin{eqnarray}\label{evol-eqn}
\rho_{de}= \rho_{de0}\, \left(\frac{a}{a_0}\right)^{-3} \, \exp\left(-3 \int_{a_0}^{a}\frac{w_{de}}{a'}\,  da'\right)
\end{eqnarray}
where $\rho_{de,0}$ is the present value of $\rho_{de}$, $a_0$ is the present value of the scale factor and $1+z = a_0/a$. In the rest of our analysis we shall consider $a_0 =1$. 
Now, from eq. (\ref{evol-eqn}) it is evident that if $w_{de}$ is specified, one can understand the possible evolution of the DE in the FLRW universe.
In our study we mainly concentrate on three dynamical DE models, namely (i) the CPL
parametrization \cite{cpl1,cpl2}, (iii) the logarithmic parametrization \cite{Efstathiou:1999tm}, and the (ii) JBP parametrization
\cite{JBP}. In what follows, we specify the basic equations that describe the evolution and dynamics of the dark energy components under such parametrizations.

\subsection{Chevallier-Polarski-Linder (CPL) model}

A simple parameterization was introduced in \cite{cpl1,cpl2} to investigate the possible dynamical aspects of DE

\begin{eqnarray}\label{cpl}
w_{de} = w_0 + w_1 \left(1-a\right),
\end{eqnarray}
where $w_0$, $w_1$ are the free parameters of the model and
physically $w_0= w_{de}(z= 0)$, i.e., it is the present value of the DE density and
$w_1 = dw/dz|_{(z= 0)}$.  The same notation is maintained for the next two models. Now, for this parametrization the DE evolution
is described by

\begin{eqnarray}\label{DE-cpl}
\rho_{de} = \rho_{de,0}\, a^{-3 \left(1+w_0 + w_1 \right)}\, \exp\left[-\,3 w_1 \left(1-a\right)\right],
\end{eqnarray}
where $\rho_{de,0}$, is the current value of DE density.

\subsection{The logarithmic model}

Let us recall another parametrization intrdouced by Efstathiou \cite{Efstathiou:1999tm} in which the equation of state (EoS) is characterized by a logarithmic law

\begin{eqnarray}\label{log}
w_{de} = w_0 - w_1 \ln a,
\end{eqnarray}
and the DE for this EoS evolves as
\begin{eqnarray}\label{log1}
\rho_{de} = \rho_{de,0} a^{-3(1+w_0)}\, \exp\left[\frac{3w_1}{2}\, \left(\ln a\right)^2 \right],
\end{eqnarray}
and commonly this parametrization is known as logarithmic parametrization.

\subsection{Jassal-Bagla-Padmanabhan (JBP) model}

Let us introduce another DE parametrization

\begin{eqnarray}\label{jbp}
	w_{de} = w_0+ w_1 a\left(1- a \right),
\end{eqnarray}
where the DE evolves as
\begin{eqnarray}
	\rho_{de}= \rho_{de,0}\, a^{-3 (1+w_0)}\, \exp\left(\frac{3w_1}{2}\,\left( a-1 \right)^2 \right)
\end{eqnarray}
This parametrization is known as the JBP parametrization.

\subsection{Linear perturbations}

Let us now review the linear perturbation equations.
The most general scalar mode perturbation is defined by the following metric \cite{Mukhanov, Ma:1995ey, Malik}

\begin{eqnarray}
ds^2 = -(1+ 2 \phi) dt^2 + 2 a \partial_i B dt dx + \nonumber \\
a^2 [(1-2 \psi) \delta_{ij} + 2 \partial_i \partial_j E] dx^i dx^j.
\end{eqnarray}

Here, we follow \cite{Ma:1995ey}, and let us adopt the synchronous gauge, i.e.
$\phi = B = 0$, $\psi = \eta$, and $k^2 E = - h/2 - 3 \eta$. The energy and momentum conservation
equations for the $i^{th}$ component of the fluid in the synchronous gauge are given by
\begin{align}
\delta'_{i} & = - (1+ w_i)\, \left(\theta_{i}+ \frac{h'}{2}\right) - 3 \mathcal{H} \left( \frac{\delta P_i}{\delta \rho_i} - w_i \right) \delta_i\,,\\
\theta'_i & = - \mathcal{H} (1- 3w_i)\theta_i- \frac{w'_i}{1+w_i} \theta_i + \frac{\delta P_i/\delta \rho_i}{1+w_i}\, k^2\, \delta_i- k^2 \, \sigma
\end{align}
where the prime denotes the derivative with respect to conformal time, $\mathcal{H}$ is the conformal Hubble function, the quantities $\sigma$, $\delta_i$, $\theta_i$, are respectively the shear, density perturbation,
velocity pertubation. The DE perturbations can be written as
\begin{align}
\delta'_{de} & = - (1+ w_{de})\, \left(\theta_{de}+ \frac{h'}{2}\right) - 3\mathcal{H}w'_{de}\frac{\theta_{de}}{k^2} \nonumber \\
&- 3 \mathcal{H} \left(c^2_s - w_{de} \right) \left[ \delta_{de}+3\mathcal{H}(1+w_{de})\frac{\theta_{de}}{k^2} \right]\,,\\
\theta'_{de} & = - \mathcal{H} (1- 3c^2_s)\theta_{de} +
\frac{c^2_s}{1+w_{de}}\, k^2\, \delta_{de}
\end{align}
where for simplicity we considered $\sigma=0$. Here, $c^2_s$, is the physical sound speed in the rest frame.
In order to avoid the unphysical sound speed we assume $c^2_s = 1$.
Since the baryons, dark matter, photons and massive neutrinos are conserved independently, thus, the perturbation equations for each component
follow the standard evolution described in \cite{Mukhanov, Ma:1995ey, Malik}.

\begingroup
\begin{table*}[ht]
	%\begin{center}
	\begin{tabular}{|r|c|c|c|c|c|c|c|c|c|c|}
		\hline \small
		\footnotesize Parameters & \multicolumn{3}{c|}{\footnotesize CPL model}\\
		\cline{2-3}
		\cline{2-3}
		\cline{3-4}
		&\footnotesize Normal Hierarchy &\footnotesize Inverted Hierarchy &\footnotesize Degenerate Hierarchy \\
		\hline
		%\hline
		\footnotesize $\Omega_{dm} h^2$ &{\footnotesize $0.1181_{-    0.0012-0.0023}^{+ 0.0012+ 0.0024}$}&{\footnotesize $0.1182_{-    0.0012- 0.0025}^{+0.0012+ 0.0024}$}&{\footnotesize $0.1182_{-0.0012-0.0024}^{+0.0012+0.0025}$}\\
		
		\footnotesize $\Omega_b h^2$ &{\footnotesize $0.0223_{- 0.0001- 0.0003}^{+ 0.0001+ 0.0003}$}&{\footnotesize $0.0223_{- 0.0001-    0.0003}^{+    0.0001+    0.0003}$}&{\footnotesize $0.0223_{-0.0001-0.0003}^{+0.0001+0.0003}$}\\
		
		\footnotesize $100\theta_{MC}$ &{\footnotesize $1.0408_{- 0.0003- 0.0006}^{+ 0.0003+  0.0006}$}&{\footnotesize $1.0408_{- 0.0003- 0.0006}^{+ 0.0003+    0.0006}$}&{\footnotesize $1.0408_{-0.0003-0.0006}^{+0.0003+0.0006}$}\\
		
		\footnotesize $\tau_{\rm reio}$ &{\footnotesize $0.0680_{- 0.0168- 0.0348}^{+ 0.0170 + 0.0332 }$}&{\footnotesize $0.0706_{- 0.0171-    0.0350}^{+ 0.0188+ 0.0334}$}&{\footnotesize $0.0680_{-0.0178-0.0351}^{+0.0181+0.0354}$}\\
		
		\footnotesize $n_s$ &{\footnotesize $0.9682_{-0.0043- 0.0083}^{+ 0.0043+ 0.0085 }$}&{\footnotesize $0.9678_{-    0.0046- 0.0082}^{+    0.0042+    0.0087}$}&{\footnotesize $0.9677_{-0.0042-0.0083}^{+0.0042+0.0084}$}\\

	\footnotesize ${\rm{ln}}(10^{10} A_s)$ &{\footnotesize $3.0655_{- 0.0325- 0.0670}^{+ 0.0328 + 0.0650 }$}&{\footnotesize $3.0706_{- 0.0341- 0.0683}^{+  0.0342+  0.0649}$}&{\footnotesize $3.0658_{-0.0345-0.0682}^{+0.0350+0.0685}$}\\
		
	\footnotesize $w_0$ &{\footnotesize $-0.9414_{- 0.1172-  0.2016}^{+ 0.1007 +0.2062 }$}&{\footnotesize $-0.9323_{- 0.1298- 0.2056}^{+    0.0957+    0.2187}$}&{\footnotesize $-0.9164_{-0.1135-0.2080}^{+0.1001+0.2127}$}\\
		
	\footnotesize $w_1$ &{\footnotesize $-0.4303_{- 0.3401- 0.8888}^{+ 0.5402 +0.8084}$}&{\footnotesize $-0.5230_{-    0.3210-    0.9270}^{+    0.5463+    0.8123}$}&{\footnotesize $-0.5564_{-0.3626-1.0445}^{+0.5824+0.9207}$}\\
		
\footnotesize $\Sigma m_\nu$ &{\footnotesize $<0.3538$ (95\% CL)}&{\footnotesize $<0.3686$ (95\% CL)}&{\footnotesize $<0.4272$ (95\% CL)}\\
                \hline

 \footnotesize $\sigma_8$ &{\footnotesize $0.8082_{- 0.0152- 0.0390}^{+    0.0211+    0.0351}$ }&{\footnotesize $0.8047_{-    0.0160-    0.0351}^{+    0.0178+    0.0350}$}&{\footnotesize $0.8068_{-0.0166-0.0431}^{+0.0236+0.0388}$}\\
		
		\footnotesize $H_0$ &{\footnotesize $68.5265_{- 0.9353- 1.6824}^{+    0.8406+    1.7887}$}&{\footnotesize $68.4954_{- 0.8758- 1.7443}^{+    0.8675+    1.7329}$}&{\footnotesize $68.4419_{-0.8708-1.6711}^{+0.8776+1.7318}$}\\

\footnotesize $\Omega_\nu h^2$ &{\footnotesize $0.0017_{- 0.0012- 0.0014}^{+ 0.0004+    0.0022}$}&{\footnotesize $0.0022_{- 0.0011-    0.0014}^{+    0.0004+    0.0019}$}&{\footnotesize $0.0019_{-0.0019-0.0022}^{+0.0007+0.0029}$}\\

\footnotesize $\Omega_{m0}$ &{\footnotesize $0.3029_{- 0.0087-    0.0169}^{+ 0.0086+    0.0178}$}&{\footnotesize $0.3042_{- 0.0093-    0.0162}^{+ 0.0084+  0.0175}$}&{\footnotesize $0.3042_{-0.0090-0.0172}^{+0.0089+0.0183}$}\\
\hline
\footnotesize $\chi^2_{min}$ &{\footnotesize $13720.722$}&{\footnotesize $ 13722.104$}&{\footnotesize $13720.610$}\\
        	\hline

	\end{tabular}
	\caption{The table summarizes the observational constraints on the free parameters and the derived parameters
	($\sigma_8, H_0, \Omega_{m0}, \Omega_{\nu} h^2$) of the CPL parametrization for three different neutrino mass hierarchies,
	namely normal hierarchy, inverted hierarchy and the degenerate hierarchy,
	using the observational data combinations CMB $+$ SNIa $+$ BAO $+$ RSD $+$ WL $+$ CC $+$ $H_0$. Mean values of the parameters
	are displayed at 1$\sigma$ (68\%) and 2$\sigma$ (95\%) errors. The parameter $H_0$ is in the units of km/s/Mpc,
	$\sum m_{\nu}$ is in the units of eV with 2$\sigma$ upper bound, and $\Omega_{m0}= \Omega_{dm0}+\Omega_{b0}$.}
	\label{tab:CPL}
%\end{center}
\end{table*}
\endgroup

\begingroup \squeezetable
\begin{table*}[ht]
%\begin{center}
\begin{tabular}{|r|c|c|c|c|c|c|c|c|c|c|}
\hline \small
 \footnotesize Parameters & \multicolumn{3}{c|}{\footnotesize Logarithmic Model}\\
\cline{2-3}
\cline{2-3}
\cline{3-4}
&\footnotesize Normal Hierarchy &\footnotesize Inverted Hierarchy &\footnotesize Degenerate Hierarchy  \\
\hline
%\hline
\footnotesize $\Omega_{dm} h^2$ &{\footnotesize $0.1184_{- 0.0012- 0.0025}^{+    0.0012+    0.0023}$}&{\footnotesize $0.1184_{-    0.0012-    0.0025}^{+    0.0012+    0.0024}$}&{\footnotesize $0.1184_{-    0.0012-    0.0025}^{+    0.0012+    0.0025}$}\\

\footnotesize $\Omega_b h^2$ &{\footnotesize $0.0223_{- 0.0001- 0.0003}^{+    0.0001+    0.0003}$}&{\footnotesize $0.0223_{-    0.0002-    0.0003}^{+    0.0002+    0.0003}$}&{\footnotesize $0.0223_{-    0.0001-    0.0003}^{+    0.0002+    0.0003}$}\\

\footnotesize $100\theta_{MC}$ &{\footnotesize $1.0408_{- 0.0003- 0.0007}^{+    0.0003+    0.0006}$}&{\footnotesize $1.0408_{- 0.0003-  0.0006}^{+    0.0003+    0.0006}$}&{\footnotesize $1.0408_{-    0.0003-    0.0006}^{+    0.0003+    0.0006}$}\\

\footnotesize $\tau_{\rm reio}$ &{\footnotesize $0.0685_{- 0.0169-    0.0332}^{+    0.0171+    0.0338}$}&{\footnotesize $0.0683_{-    0.0178-    0.0351}^{+    0.0181+    0.0352}$}&{\footnotesize $0.0669_{-    0.0178-    0.0355}^{+    0.0181+    0.0355}$}\\

\footnotesize $n_s$ &{\footnotesize $0.9676_{- 0.0044 -0.0086}^{+    0.0045+    0.0087}$}&{\footnotesize $0.9674_{-    0.0044-    0.0085}^{+    0.0044+    0.0085}$}&{\footnotesize $0.9674_{-    0.0042-    0.0086}^{+    0.0043+    0.0085}$}\\

\footnotesize ${\rm{ln}}(10^{10} A_s)$ &{\footnotesize $3.0670_{- 0.0332- 0.0643}^{+    0.0334+    0.0650}$}&{\footnotesize $3.0666_{-    0.0339-    0.0669}^{+    0.0353+    0.0689}$ }&{\footnotesize $3.0642_{-    0.0343-    0.0685}^{+    0.0350+    0.0678}$}\\

\footnotesize $w_0$ &{\footnotesize $-0.9074_{- 0.0962- 0.1628}^{+    0.0728+    0.1688}$}&{\footnotesize $-0.9147_{-    0.1035-    0.1707}^{+    0.0763+    0.1789}$}&{\footnotesize $-0.9143_{-    0.0958-    0.1563}^{+    0.0683+    0.1765}$}\\

\footnotesize $w_1$ &{\footnotesize $-0.4727_{-  0.1479- 0.6136}^{+    0.4055+    0.4727}$}&{\footnotesize $-0.4904_{-    0.1425-    0.6493}^{+    0.4404+    0.4904}$}&{\footnotesize $-0.4358_{-    0.0986-    0.6459}^{+    0.4358+    0.4358}$}\\

\footnotesize $\Sigma m_\nu$ &{\footnotesize $<0.412$ (95\% CL)}&{\footnotesize $<0.428$ (95\% CL)}&{\footnotesize $<0.425$ (95\% CL)}\\
\hline

\footnotesize $\sigma_8$ &{\footnotesize $0.8072_{- 0.0149- 0.0406}^{+    0.0208+    0.0378}$ }&{\footnotesize $0.8036_{- 0.0154- 0.0395}^{+    0.0207+    0.0372}$}&{\footnotesize $0.8092_{- 0.0176- 0.0455}^{+  0.0242+    0.0403}$}\\

\footnotesize $H_0$ &{\footnotesize $68.3616_{-    0.8735-    1.6450}^{+    0.8455+    1.6910}$}&{\footnotesize $68.4264_{-    0.8408-    1.6902}^{+    0.8587+    1.6949}$}&{\footnotesize $68.4151_{-    0.8734-    1.6691}^{+    0.8364+    1.7241}$}\\

\footnotesize $\Omega_\nu h^2$ &{\footnotesize $0.0020_{-    0.0014-    0.0018-    0.0013}^{+    0.0004+    0.0026}$}&{\footnotesize $0.0023_{-    0.0013-    0.0016}^{+    0.0004+    0.0024}$}&{\footnotesize $0.0018_{-    0.0018-    0.0021}^{+    0.0007+    0.0030}$}\\

\footnotesize $\Omega_{m0}$ &{\footnotesize $0.3054_{-    0.0091-    0.0159}^{+    0.0083+    0.0178}$}&{\footnotesize $0.3055_{-    0.0094-    0.0172}^{+    0.0084+    0.0176}$}&{\footnotesize $0.3045_{-    0.0097-    0.0163}^{+    0.0081+    0.0181}$}\\

\hline
\footnotesize $\chi^2_{min}$ &{\footnotesize $13723.006$}&{\footnotesize $ 13721.892$}&{\footnotesize $13722.96$}\\
\hline

\end{tabular}
\caption{The table summarizes the observational constraints on the free parameters and the derived parameters 
	($\sigma_8, H_0, \Omega_{m0}, \Omega_{\nu} h^2$) of the logarithmic parametrization for three different neutrino mass hierarchies, 
namely the normal hierarchy, inverted hierarchy and the degenerate hierarchy,
	using the observational data combinations CMB $+$ SNIa $+$ BAO $+$ RSD $+$ WL $+$ CC $+$ $H_0$. Mean values of the parameters
	are displayed at 1$\sigma$ (68\%) and 2$\sigma$ (95\%) errors. The parameter $H_0$ is in the units of km/s/Mpc, 
	$\sum m_{\nu}$ is in the units of eV with 2$\sigma$ upper bound, and $\Omega_{m0}= \Omega_{dm0}+\Omega_{b0}$.}
\label{tab:log}
%\end{center}
\end{table*}
\endgroup

\begin{figure*}[!htbp]
	\includegraphics[width=12cm,height=10cm]{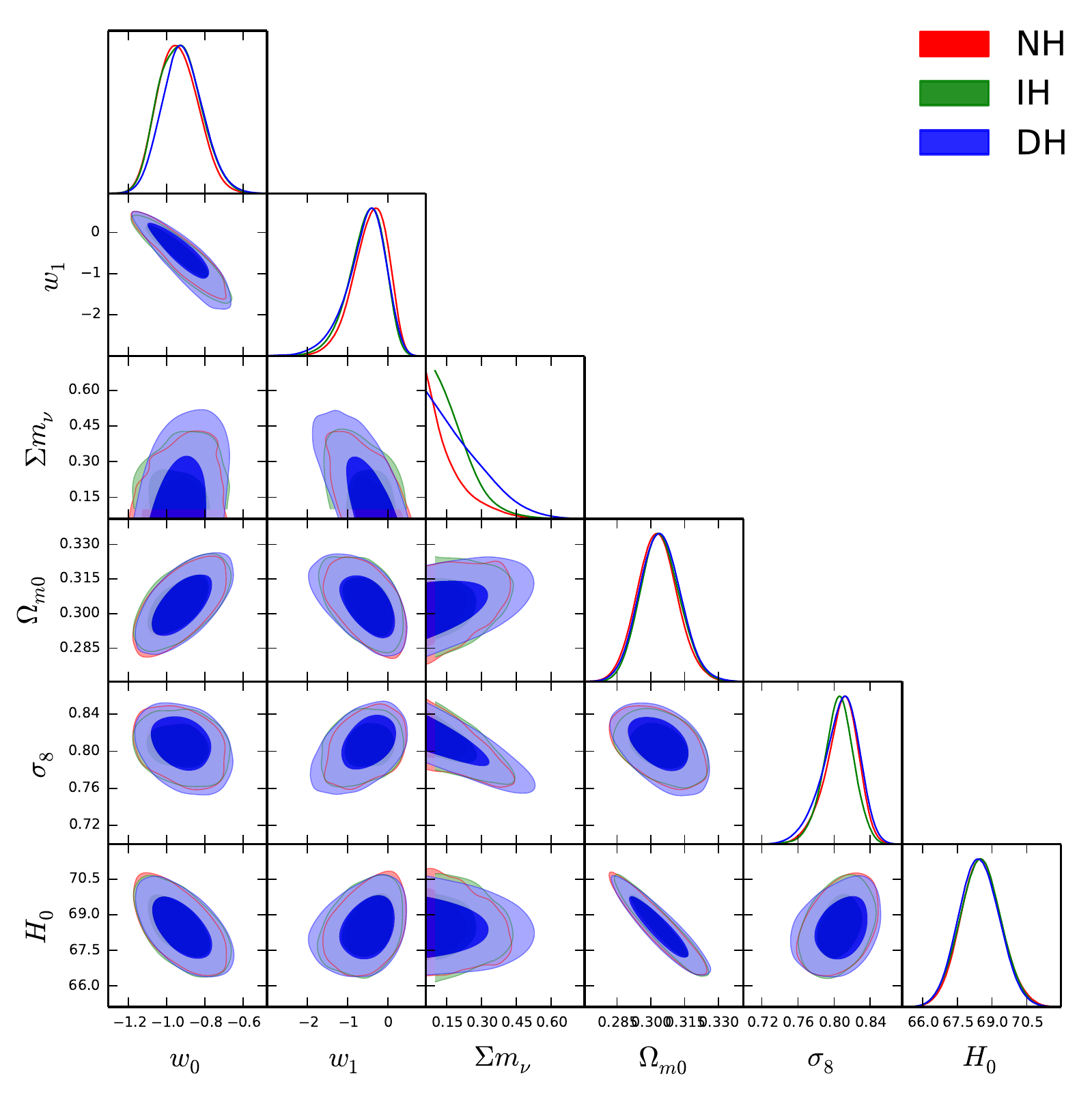}
	\caption{68\% and 95\% confidence-level contour plots for some selected parameters of the CPL parametrization
	considering three different neutrino mass hierarchies, namley the NH, IH, DH using CMB $+$ SNIa $+$ BAO $+$ RSD $+$ WL $+$ CC $+$ $H_0$ data set. }
	\label{fig:contour-cpl}
\end{figure*}

\section{Data set and Methodology}
\label{sec-data}

In what follows, we briefly describe the observational data sets used in this work.

\begin{enumerate}

\item \textit{Cosmic Microwave Background data (CMB):}
We use CMB data from the Planck 2015 measurements
\cite{ref:Planck2015-1, ref:Planck2015-2}, where we combine the likelihoods $C^{TT}_l$, $C^{EE}_l$, $C^{TE}_l$
in addition with low$-l$ polarization $C^{TE}_l+C^{EE}_l+C^{BB}_l$,
which notationally is same with ``Planck TT, EE, TE + lowTEB'' of Ref. \cite{Ade:2015xua}.

\item \textit{Supernovae Type Ia (SNIa):} We take the latest joint light curves (JLA) sample \cite{Betoule:2014frx} containing 740 SNIa in the redshift range $z\in[0.01, 1.30]$.

\item \textit{Baryon acoustic oscillations (BAO) distance measurements:} For the BAO data we use the estimated ratio $r_s/D_V$ as a `standard ruler' in which $r_s$ is the comoving sound horizon at the baryon drag epoch and $D_V$ is the effective distance determined by the angular diameter distance $D_A$ and Hubble parameter $H$ as $D_V(z)=\left[(1+z)^2D_A(a)^2\frac{z}{H(z)}\right]^{1/3}$. We consider three different measurements as, $r_s(z_d)/D_V(z=0.106)=0.336\pm0.015$ from 6-degree Field Galaxy Redshift Survey (6dFGRS) data \cite{Beutler:2011hx}, $r_s(z_d)/D_V(z=0.35)=0.1126\pm0.0022$ from Sloan Digital
Sky Survey Data Release 7 (SDSS DR7) data \cite{Padmanabhan:2012hf}, and finally $r_s(z_d)/D_V(z=0.57)=0.0732\pm0.0012$ from the SDSS DR9 \cite{Manera:2012sc}.

\item \textit{Redshift space distortion (RSD):} We use RSD data from different observational surveys from 2dFGRS \cite{Percival:2004fs}, the
WiggleZ \cite{Blake:2011rj}, the SDSS LRG \cite{Samushia:2011cs}, the BOSS CMASS \cite{Reid:2012sw}, the 6dFGRS \cite{Beutler:2012px}, and the VIPERS \cite{delaTorre:2013rpa}. The measured values of the RSD data can be found in Table I of Ref. \cite{Yang:2014hea}.

\item \textit{Weak lensing (WL) data:} We consider the weak gravitational lensing
data from blue galaxy sample compliled from  Canada$-$France$-$Hawaii Telescope Lensing Survey
(CFHTLenS) \cite{Heymans:2013fya,Asgari:2016xuw} for our analysis.

\item \textit{Cosmic chronometers (CC) plus the local value of the Hubble parameter (CC $+$ $H_0$):}
We employ the recently released cosmic chronometers data comprising $30$ measurements of the Hubble parameter
in the redshift interval $0 < z< 2$ \cite{Moresco:2016mzx}.
Additionally we use the local value of the Hubble parameter
yielding $H_0= 73.02 \pm 1.79$  km/s/Mpc with 2.4\% precision as reported recently in \cite{Riess:2016jrr}.

\end{enumerate}

We modified the publicly available CosmoMc code \cite{cosmomc} to obtain the Markov Chain Monte Carlo samples using uniform priors 
on the free parameters

\begin{eqnarray}
\label{P1}
P = \{\omega_{b}, \, \omega_{dm},  \, 100\theta_{MC}, \, \ln10^{10}A_{s}, \, \nonumber \\
n_s, \, \tau_{\rm reio}, \, \sum m_{\nu}, \, w_0, \, w_1 \},
\end{eqnarray}
where $\omega_b= \Omega_b h^2$, $\omega_{dm}= \Omega_{dm} h^2$ are respectively the baryon density and the cold dark matter
density, $\theta_{MC}$ is the approximation to the angular size of sound horizon at last scattering, $A_s$ is defined to be 
the amplitude of initial power spectrum, $n_s$ is the spectral index, $\tau _{\rm reio}$ is the optical depth due to reionization;
$\sum m_{\nu}$ is the total neutrino mass; $w_0$, $w_1$ are the model parameters which have been defined previously. 
The priors used for the model parameters are: $\omega_b\in[0.005,0.1]$, $\omega_{m}\in[0.01,0.99]$, $100\theta_{MC}\in[0.5,10]$,
$\ln(10^{10}A_s)\in[2.4,4]$, $n_s\in[0.5,1.5]$, $\tau_{\rm reio}\in[0.01,0.8]$, $w_0\in[-2,0]$ and $w_1\in[-3,3]$ 
for CPL and JBP model while for logarithmic parameterization $w_0\in[-2,0]$ and $w_1\in[-3,0]$.
In order to constrain the free parameters of the models, we consider three dynamical DE models in normal hierarchy (NH),
inverted hierarchy (IH), degenerate hierarchy (DH) with a minimum sums of neutrino mass to be $0.06$ eV, $0.1$ eV, and $0.0$ eV,
respectively, on the three species of active neutrinos. In what follows, for the present analysis, we have considered the PPF
approximation \cite{ppf,ppf2} for three DE models.
In particular, for the logarithmic model in eq. (\ref{log}) if the prior on $w_1$ is considered to be positive, i.e. $w_1 > 0$, then
when the function $w(z)$ is evaluated at very early times, it has a positive divergence implying $w_{de} > -1/3$  
in the radiation era. Therefore, to avoid such problems, we have fixed the prior as $w_1 \leq 0$ during 
the statistical analysis. For all dynamical DE models we run the Monte Carlo Markov Chains until 
the parameters converge to a parameter according to the Gelman-Rubin criteria $R - 1 < 0.01$ \cite{Gelman-Rubin}.

\section{Results of the analysis}
\label{results}

Let us summarize the main observational results extracted from the dynamical dark energy models in presence of massive neutrinos at
three different hierarchies, namely the NH, IH, and DH, using the combined observational data,
CMB $+$ SNIa $+$ BAO $+$ RSD $+$ WL $+$ CC $+$ $H_0$, described in Section \ref{sec-data}.
In Tables \ref{tab:CPL}, \ref{tab:log}, \ref{tab:JBP} we summarize the main results of the statistical analysis for CPL, 
logarithmic, and JBP models, respectively.

\begingroup
\squeezetable
\begin{table*}[ht]
	%\begin{center}
	\begin{tabular}{|r|c|c|c|c|c|c|c|c|c|c|}
		\hline \small
		\footnotesize Parameters & \multicolumn{3}{c|}{\footnotesize JBP model}\\
		\cline{2-3}
		\cline{2-3}
		\cline{3-4}
		&\footnotesize Normal Hierarchy &\footnotesize Inverted Hierarchy &\footnotesize Degenerate Hierarchy \\
		\hline
		%\hline
		\footnotesize $\Omega_{dm} h^2$ &{\footnotesize $0.1179_{- 0.0013-    0.0025}^{+ 0.0012+ 0.0024}$}&{\footnotesize $0.1179_{-    0.0012-    0.0024}^{+    0.0012+    0.0023}$}&{\footnotesize $0.1180_{-    0.0012-    0.0025}^{+    0.0012+    0.0024}$}\\
		
		\footnotesize $\Omega_b h^2$ &{\footnotesize $0.0223_{-    0.0001-    0.0003}^{+    0.0002+    0.0003}$}&{\footnotesize $0.0223_{-    0.0001-    0.0003}^{+    0.0001+    0.0003}$}&{\footnotesize $0.0223_{-    0.0001-    0.0003}^{+    0.0001+    0.00029}$}\\
		
		\footnotesize $100\theta_{MC}$ &{\footnotesize $1.0409_{-    0.0003-    0.0006}^{+    0.0003+    0.0006}$}&{\footnotesize $1.0408_{-    0.0003-    0.0006}^{+    0.0003+    0.0006}$}&{\footnotesize $1.0409_{-    0.0003-    0.0006}^{+    0.0003+    0.0006}$}\\
		
\footnotesize $\tau_{\rm reio}$ &{\footnotesize $0.0703_{-    0.0182-    0.0341}^{+    0.0171+    0.0346}$}&{\footnotesize $0.0708_{-    0.0170-    0.0334}^{+    0.0171+    0.0337}$}&{\footnotesize $0.0680_{-    0.0177-    0.0340}^{+    0.0178+    0.0348}$}\\
		
\footnotesize $n_s$ &{\footnotesize $0.9689_{-    0.0047-    0.0085}^{+    0.0041+    0.0088}$}&{\footnotesize $0.9686_{-    0.0041-    0.0080}^{+    0.0041+    0.0082}$}&{\footnotesize $0.9686_{-    0.0041-    0.0086}^{+    0.0042+    0.0084}$}\\

		\footnotesize ${\rm{ln}}(10^{10} A_s)$ &{\footnotesize $3.0701_{-    0.0331-    0.0650}^{+    0.0332+    0.0665}$}&{\footnotesize $3.0706_{-    0.0328-    0.0650}^{+    0.0328+    0.0655}$ }&{\footnotesize $3.0658_{-    0.0343-    0.0658}^{+    0.0342+    0.0673}$}\\
		
		\footnotesize $w_0$ &{\footnotesize $-0.9628_{-    0.1846-    0.2900}^{+    0.1454+    0.3222}$}&{\footnotesize $-0.9329_{-    0.1572-    0.2767}^{+    0.1443+    0.2802}$}&{\footnotesize $-0.9658_{- 0.1344- 0.2372}^{+ 0.1163+ 0.2602}$}\\
		
		\footnotesize $w_1$ &{\footnotesize $-0.5080_{-    0.8641-    2.0100}^{+    1.1835+    1.6937}$}&{\footnotesize $-0.7903_{-    0.8594-    1.8009}^{+    0.9842+    1.6452}$}&{\footnotesize $-0.3964_{-    0.6569-    1.6432}^{+    0.8292+    1.4567}$}\\
		
		\footnotesize $\Sigma m_\nu$ &{\footnotesize $< 0.294$ (95\% CL)}&{\footnotesize $< 0.348$ (95\% CL)}&{\footnotesize $< 0.253$ (95\% CL)}\\
		\hline
		
		\footnotesize $\sigma_8$ &{\footnotesize $0.8107_{-    0.0153-    0.0325}^{+    0.0166+    0.0303}$ }&{\footnotesize $0.8028_{-    0.0155-    0.0328}^{+    0.0179+    0.0307}$}&{\footnotesize $0.8153_{-    0.0164-    0.0345}^{+    0.0166+    0.0321}$}\\
		
		\footnotesize $H_0$ &{\footnotesize $68.6228_{-    1.0041-    1.7978}^{+    0.9276+    1.9123}$}&{\footnotesize $68.5100_{-    0.9235-    1.8080}^{+    0.9122+    1.8247}$}&{\footnotesize $68.5621_{-    0.8739-    1.6835}^{+    0.8695+    1.7740}$}\\
		
		\footnotesize $\Omega_\nu h^2$ &{\footnotesize $0.0015_{-    0.0010-    0.0012}^{+    0.0003+    0.0017}$}&{\footnotesize $0.0021_{-    0.0010-    0.0013}^{+    0.0004+    0.0017}$}&{\footnotesize $0.0011_{-    0.0010-    0.0013}^{+    0.0004+    0.0017}$}\\
		
		\footnotesize $\Omega_{m0}$ &{\footnotesize $0.3013_{-    0.0097-    0.0175}^{+    0.0089+    0.0182}$}&{\footnotesize $0.3035_{-    0.0098-    0.0168}^{+    0.0089+    0.0184}$}&{\footnotesize $0.3010_{-    0.0089-    0.0165}^{+    0.0084+    0.0165}$}\\

\hline 
\footnotesize $\chi^2_{min}$ &{\footnotesize $13721.704$}&{\footnotesize $ 13720.834$}&{\footnotesize $13719.138$}\\

		\hline
	\end{tabular}
	\caption{The table summarizes the observational constraints on the free parameters and the derived parameters ($\sigma_8, H_0, \Omega_{m0}, \Omega_{\nu} h^2$) of the JBP parametrization for three different neutrino mass hierarchies, the namely normal hierarchy, inverted hierarchy and the degenerate hierarchy, using the observational data combinations CMB $+$ SNIa $+$ BAO $+$ RSD $+$ WL $+$ CC $+$ $H_0$. Mean values of the parameters
	are displayed at 1$\sigma$ (68\%) and 2$\sigma$ (95\%) errors. The parameter $H_0$ is in the units of km/s/Mpc, $\sum m_{\nu}$ 
	is in the units of eV with 2$\sigma$ upper bound, and $\Omega_{m0}= \Omega_{dm0}+\Omega_{b0}$.}
	\label{tab:JBP}
	%\end{center}
\end{table*} 
\endgroup

\begin{figure*}[!htbp]
\includegraphics[width=12cm,height=10cm]{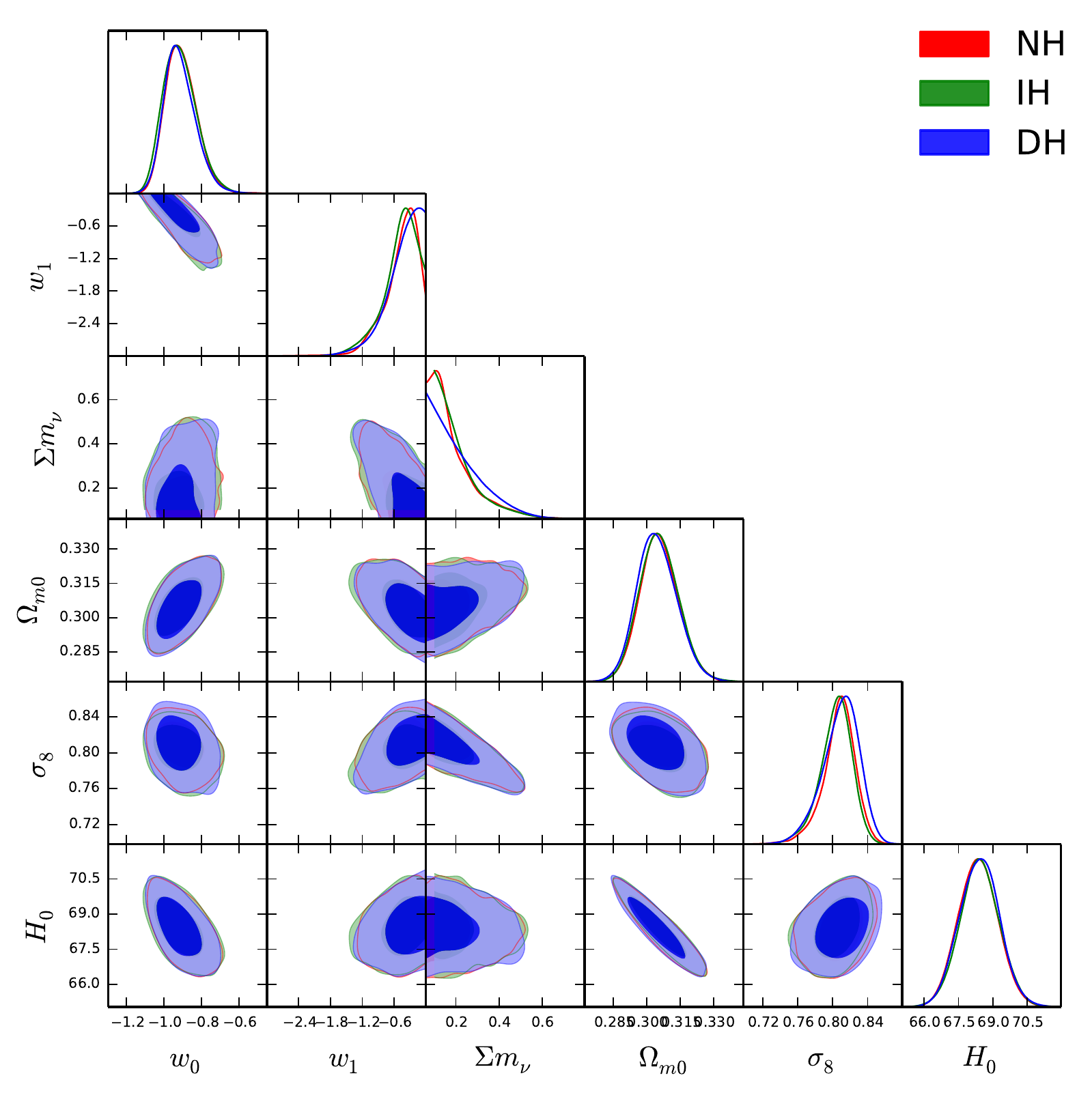}
  \caption{68\% and 95\% confidence-level contour plots for some selected parameters of the logarithmic parametrization
	considering three different neutrino mass hierarchies, namley the NH, IH, DH using
	CMB $+$ SNIa $+$ BAO $+$ RSD $+$ WL $+$ CC $+$ $H_0$.}
  \label{fig:contour-log}
\end{figure*}

\begin{figure*}[!htbp]
	\includegraphics[width=12cm,height=10cm]{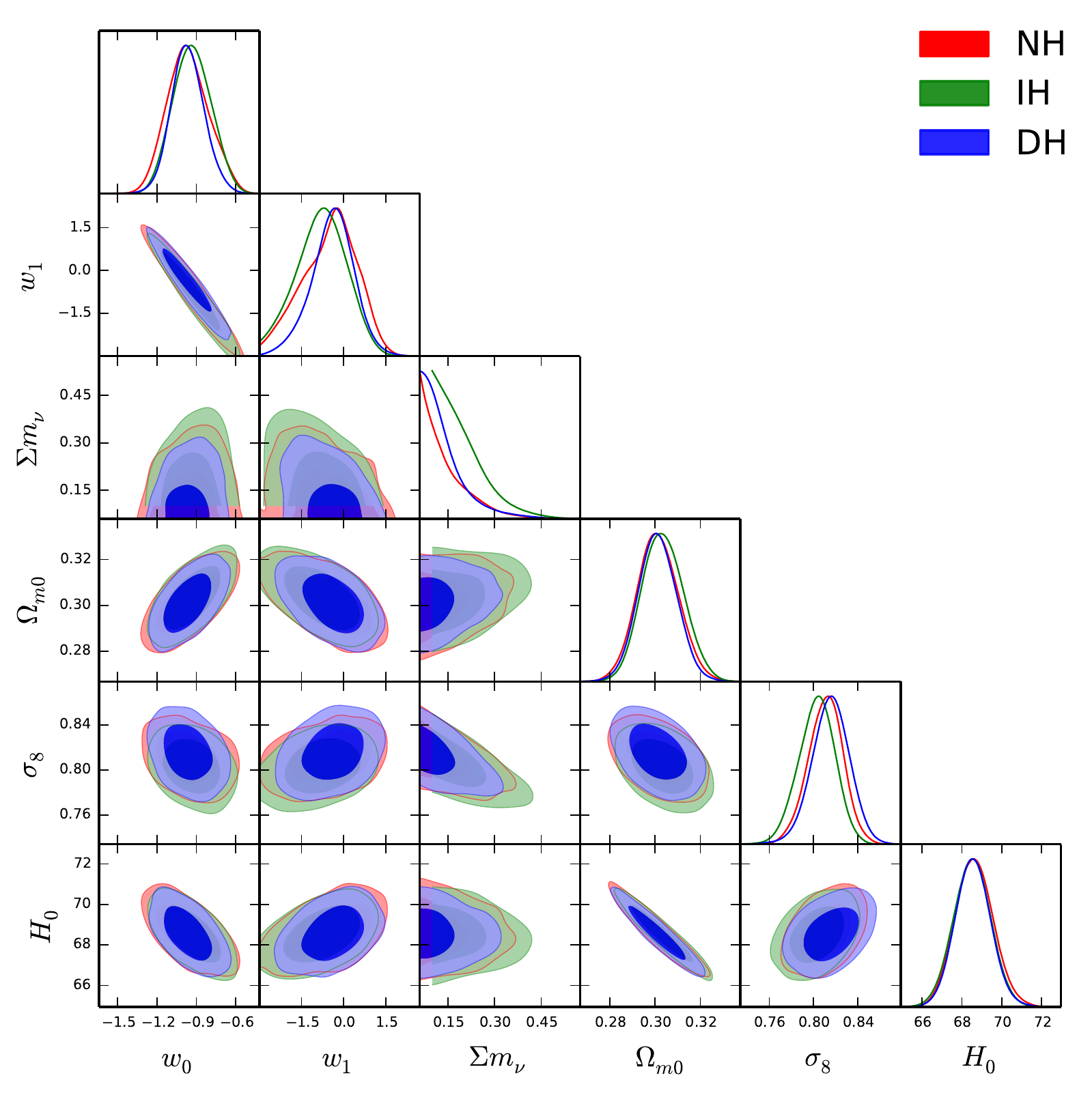}
	\caption{68\% and 95\% confidence-level contour plots for some selected parameters of the JBP model
	considering three different neutrino mass hierarchies, namley the NH, IH, DH using CMB $+$ SNIa $+$ BAO $+$ RSD $+$ WL $+$ CC $+$ $H_0$.}
	\label{fig:contour-jbp}
\end{figure*}

\begin{figure*}[!htbp]
  	\includegraphics[width=5.8cm,height=5cm]{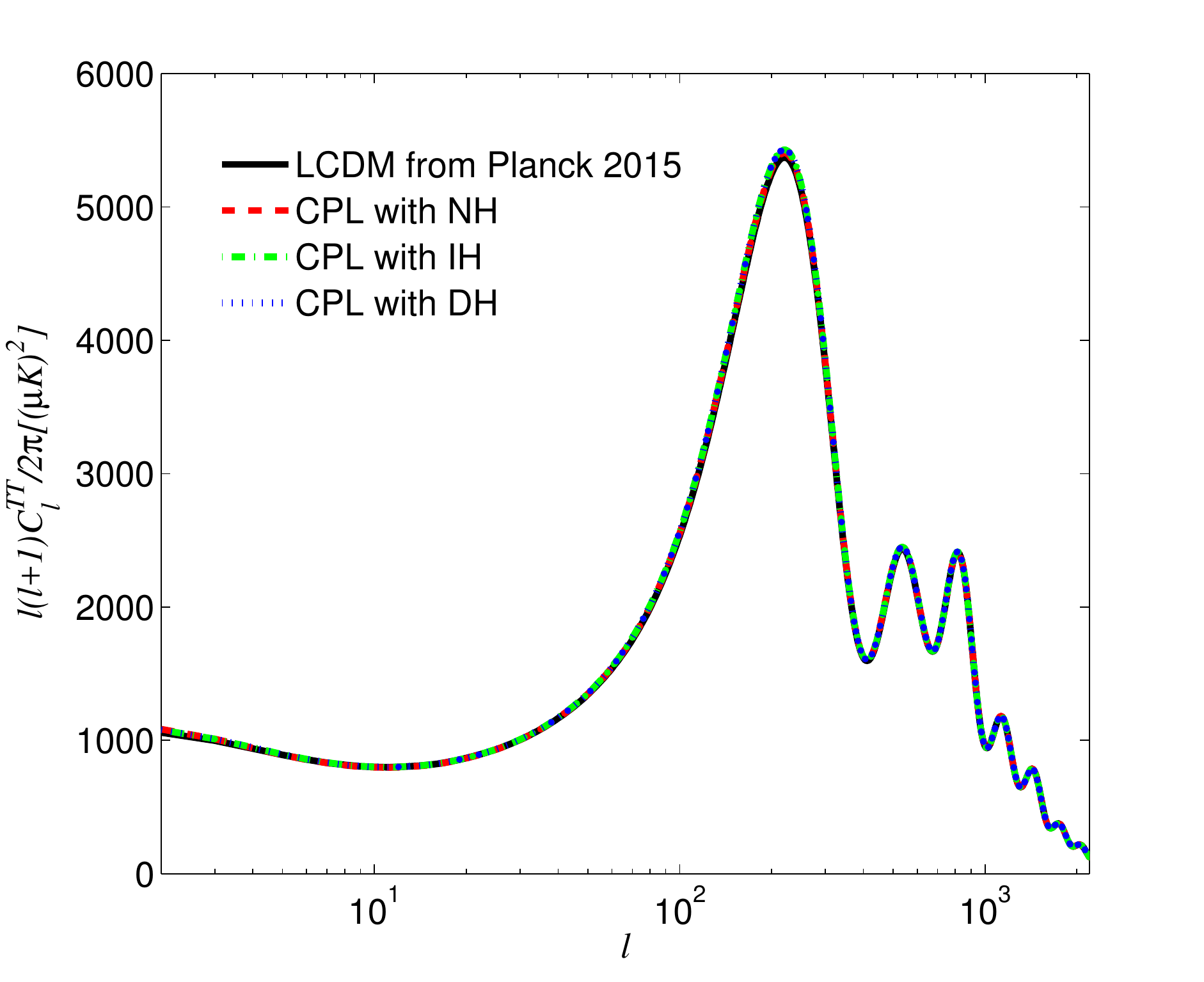}
  	\includegraphics[width=5.8cm,height=5cm]{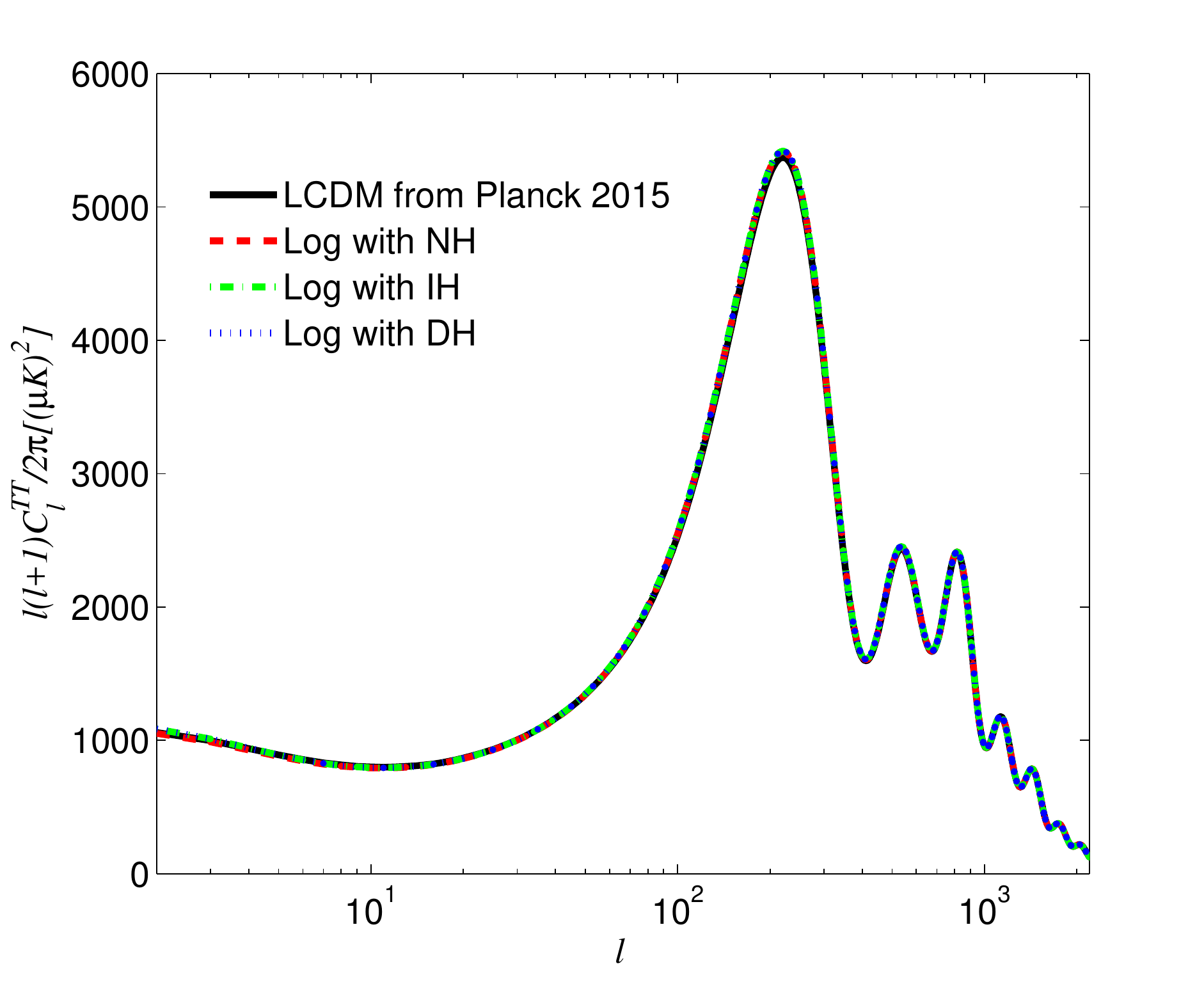}
  	\includegraphics[width=5.8cm,height=5cm]{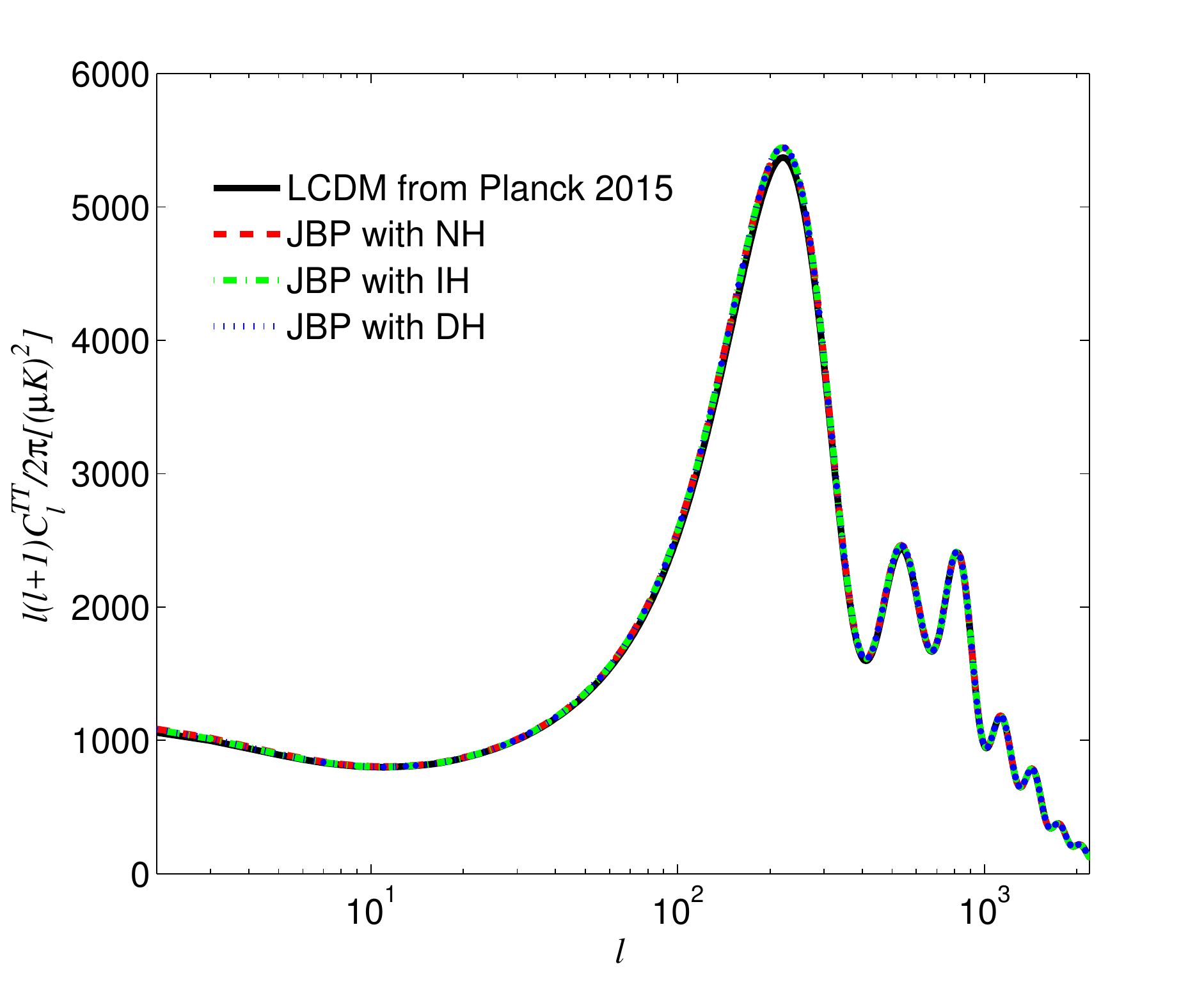}
  	\caption{The CMB TT power spectrum for CPL (left panel), logarithmic (middle panel), and JBP (right panel) models 
  	 for the three different mass hierarchy schemes, in compared to the $\Lambda$CDM TT power spectrum. We note that in all three plots the spectra for different hierarchies as well as for $\Lambda$CDM are completely indistinguishable from each other. }
  	\label{fig:compare}
\end{figure*}

\begin{figure*}[!htbp]
  	\includegraphics[width=5.45cm,height=4.5cm]{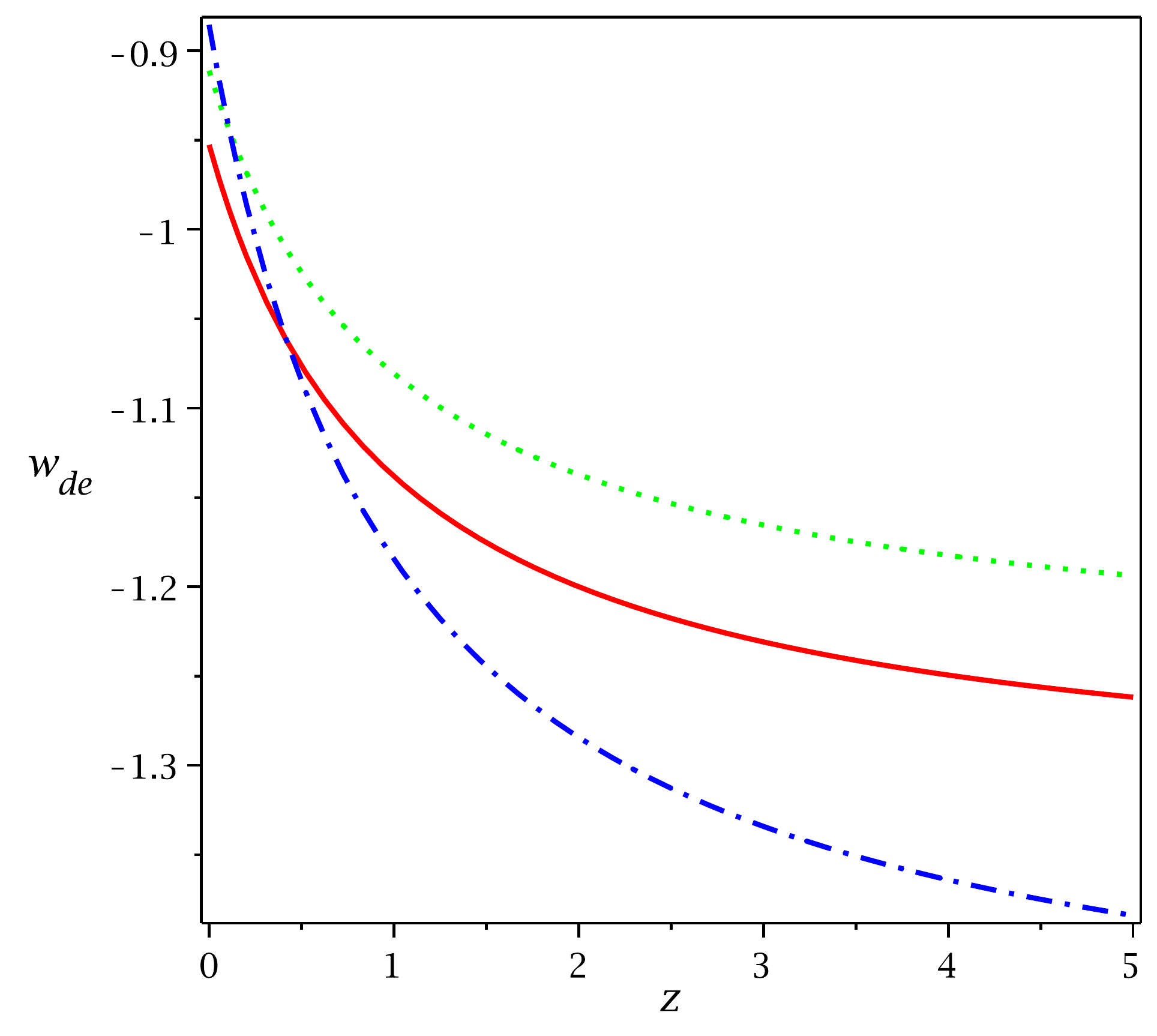}
  	\includegraphics[width=5.45cm,height=4.5cm]{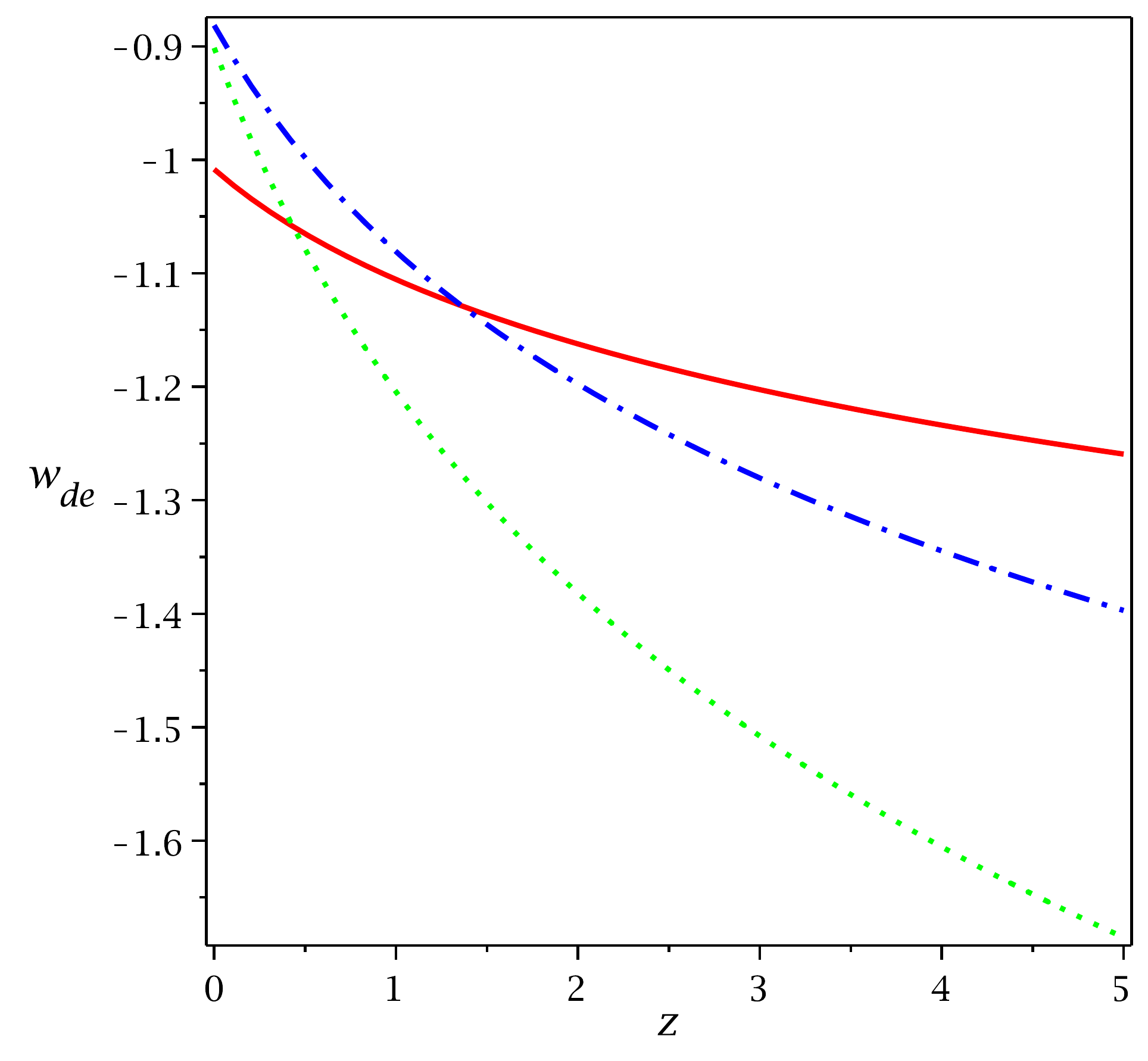}
  	\includegraphics[width=5.45cm,height=4.5cm]{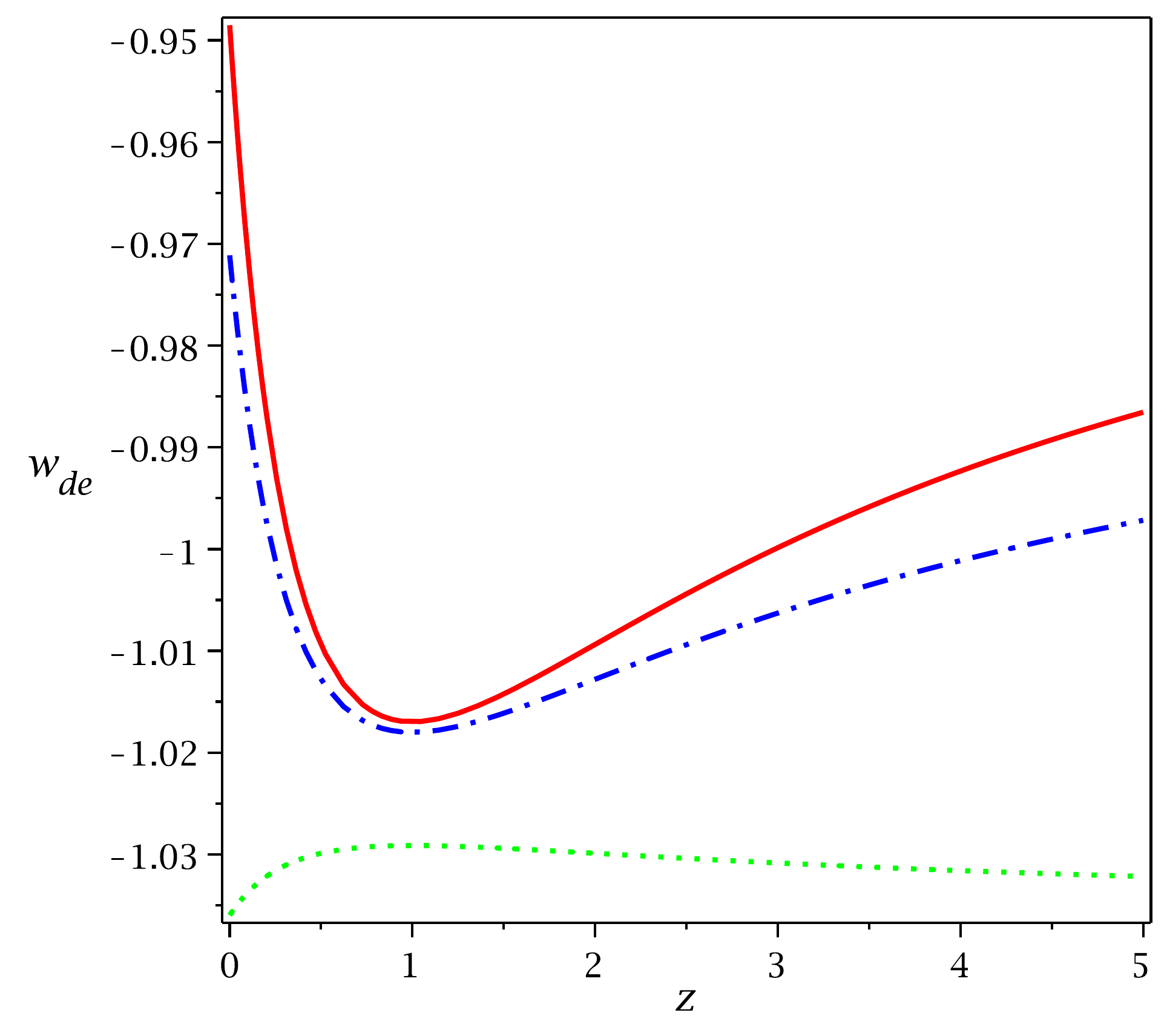}
  	\caption{Evolution of the DE EoS as a function of the redshift for CPL (left panel), logarithmic (middle panel) and JBP 
  	(right panel) parametrizations at three different neutrino mass hierarchies have been shown for the best fit values 
  	of the model parameters $w_0$ and $w_1$ using the combined analysis CMB $+$ SNIa $+$ BAO $+$ RSD $+$ WL $+$ CC $+$ $H_0$. 
  	In each plot, the solid, dot and dashdot curves stand for NH, IH and DH, respectively.}
  	\label{fig:eos-3models}
\end{figure*}

\begin{table}[ht]
%\begin{center}
\begin{tabular}{|r|c|c|c|c|c|c|c|c|c|c|}
\hline \small
Models &\footnotesize NH &\footnotesize IH &\footnotesize DH \\
\hline
%\hline
\footnotesize CPL model &{\footnotesize $ \sum m_{\nu} < 0.353 $}&{\footnotesize $ \sum m_{\nu}< 0.368 $}&{\footnotesize $\sum m_{\nu} < 0.427$}\\
\footnotesize logarithmic model &{\footnotesize $ \sum m_{\nu} < 0.412 $}&{\footnotesize $ \sum m_{\nu} < 0.428 $}&{\footnotesize $\sum m_{\nu} < 0.425 $}\\
\footnotesize JBP model &{\footnotesize $ \sum m_{\nu}< 0.294 $}&{\footnotesize $\sum m_{\nu} < 0.348 $}&{\footnotesize $\sum m_{\nu} < 0.253$}\\
\hline
\end{tabular}
\caption{Summary of the cosmological neutrino mass bound ($\sum m_{\nu}$) at 95$\%$ CL, for the three dynamical DE models considered here. 
The mass is in the units of eV.}
\label{tab:mass}
%\end{center}
\end{table}

Figure \ref{fig:contour-cpl} shows the one-dimensional marginalized distribution and 68\%, 95\% CL regions for some selected parameters of the CPL model.
We note that $\sum m_{\nu} < 0.353,\,  0.368,\,  0.444$ eV at 95\% CL, for NH, IH and DH, respectively. In general, our analysis reveals
that all three neutrino hierarchies return almost similar constraints on the baseline parameters. 
However, a slight variational effect depending on different hierarchies of the neutrino masses is observed on the DE parameter $w_1$.
 In particular we observe that the maximum variation of $w_1$ (from its best fit values)
 \footnote{Here we define the variation as $\Delta w_1 = w_1 (\mbox{at}\, \mbox{NH})- w_1 (\mbox{at}\,\mbox{IH})\,\, 
 (\mbox{or}\,\, w_1 (\mbox{at}\, \mbox{DH}))$.} is of order $\Delta w_1 \sim 0.1$. 

Figures \ref{fig:contour-log} and \ref{fig:contour-jbp} show the one-dimensional marginalized distribution and the parametric space at
68\%, 95\% CL regions for some selected parameters of the logarithmic and JBP models, respectively.
We can note that no significant variations are observed in the full parameter base of the logarithmic parameterization, 
including DE properties ($w_0$ and $w_1$). Within logarithmic model we note that $\sum m_{\nu} < 0.412,\,  0.428,\,  0.425$ eV at 95\% CL, 
for NH, IH and DH, respectively. On the other hand, in JBP model we note the significant variations on the DE parameter. In particular, on $w_1$, 
we find that $\Delta w_1 = 0.2883 \,  (\mbox{or}, - 0.1116)$ when NH is compared to IH or DH. 
Here, we note $\sum m_{\nu} < 0.294, \, 0.348, \,  0.253$ eV at 95\% CL, for NH, IH and DH, respectively.

In general, taking into account the neutrino mass splittings, for instance within the CPL model with NH scheme,
that works out to be 2 neutrinos of approximately 0.10 eV and 1 slightly heavier neutrino of 0.15 eV.
This scenario is almost degenerate. The other hierarchy schemes present practically the same upper limits on the 
neutrino mass splittings, therefore, we really should expect nearly identical results on the baseline of the model. 
The same interpretation applies to both logarithmic and JBP models. Additionally, in Figure \ref{fig:compare} we show the theoretical predictions of the angular CMB 
power spectrum temperature anisotropy for the three dynamical DE models considered in this work in comparison to the $\Lambda$CDM model. 
In those plots we have assumed the best fit values from the Tables for each respective model. Evidently one can clearly observe 
that significant variations are not observed in the behavior of the dynamical DE models when the presence of massive neutrinos 
are taken into account in the cosmological picture. In fact, our constraints are very close to the $\Lambda$CDM cosmology.

Figure \ref{fig:eos-3models} shows the quatitative evolution of the DE EoS 
considering three distinct neutrino mass schemes for CPL, logarithmic and JBP models. 
From the figure, we see that at high redshifts, the EoS for the DE parametrizations exhibit significant deviations at three 
different hierarchies. However, we notice that for $z \sim 0$, the EoS at different neutrino mass hierarchies 
for the CPL and logarithmic models become close to each other, while in the JBP parametrization, the EoS curves  
for NH and DH are similar in contrary to the EoS curve at IH.

\section{Final Remarks}
\label{conclusions}

The presence of massive neutrinos is an essential piece in the dynamics of the universe, and it is known that their
properties can correlate in different ways with other cosmological parameters.
Thus, the determination of its properties with accurate and robust way plays an important role on a particular cosmological model.

In this work we have measured the effects of massive neutrinos via three different neutrino hierarchies,
namely NH, IH, and DH, on the cosmological scenarios where DE offers a dynamical character.
We consider three well known and most used dynamical DE models represented by CPL,
logarithmic and JBP parametrizations. The models have been constrained using the
most current observational data from CMB, SNIa, BAO, RSD, WL, CC, and $H_0$.
From the combined analysis of these observational data, we provide with robustness a cosmological neutrino mass bound (see Table \ref{tab:mass}) in presence of the dynamical DE models.
Further, we have found that the fixation of different neutrino mass hierarchies does not exhibit any significant variation on the baseline of parameters of the models, 
except on the DE parameter $w_1$ in CPL and JBP parameterizations. 
But, such variations do not present statistical 
deviations from $\Lambda$CDM model. In general, we can summarize our results by concluding that independent of the dynamic nature of DE,
different choices of neutrino mass 
scheme throughout the cosmic history will not make significant 
changes on the dynamic properties of DE within each model.

\section*{Acknowledgments}

The authors thank the anonymous referee for his/her detailed comments and suggestions which improved the work significantly.
Also, the authors are grateful to Thomas Tram, Lixin Xu, and Vinicius Miranda for helpful discussions. 
W. Yang's work is supported by the National Natural
Science Foundation of China under Grant No. 11647153,
the Foundation of Education Department of Liaoning
Province in China under Grant No. L201683666, the
Youth Foundation of Liaoning Normal University under
Grant No. LS2015L003. SP is supported by the SERB$-$NPDF grant (PDF/2015/000640), Government of India.
DFM acknowledges the support from the Research Council of Norway.

\end{document}